\documentclass[a4paper,11pt]{article}
\pdfoutput=1 

\usepackage[margin=1in]{geometry}
                     
\usepackage[T1]{fontenc} 

\usepackage{slashed}
\usepackage{subfig}
\usepackage{graphicx}
\usepackage{authblk}
\usepackage{amssymb}
\usepackage{amsmath}

\newcommand{\pb}{\text{ pb}}

\newcommand{\gev}{\text{ GeV}}
\newcommand{\tev}{\text{ TeV}}

\newcommand{\mhp}{m_{H^\pm}}

\title{\boldmath{Deep-Learning in Search of Light Charged Higgs}}

\author[a]{G. K. Demir \thanks{ corresponding author,\textit{e-mail}: guleser.kalayci@deu.edu.tr}}
\author[b]{N. S{\"o}nmez}
\author[a]{H. Do{\~g}an}
\affil[a]{Department of Electrical and Electronics Engineering, Dokuz Eyl{\"u}l University, 35390, {\.I}zmir, Turkey}
\affil[b]{Department of Physics, Ege University, 35040, {\.I}zmir, Turkey}
\affil[ ]{}
\date{}

\begin{document} 
\maketitle
\begin{abstract}
In this work, we deep-learn light charged Higgs signal in top quark decays which poses difficulties due to strong W boson contamination. We construct Deep Neural Networks (DNN) with appropriate architecture and determine signal extraction efficiency by considering various features (kinematical and human engineered parameters). Results show that DNN gives better performance than the classical neural networks and has ability to find regions of high efficiency even the input features are not human-engineered. In a sense, human-engineered high-level features are offset by DNNs with different combinations of the low-level kinematical features. Additionally, it is shown that  increasing the number of processing units in DNNs does not necessarily cause an increase in efficiency due mainly to increased complexity. Our method and results can set an example of signal extraction from strong backgrounds. 
\end{abstract}

\section{Introduction}
Charged Higgs boson, frequently arising in models with more than one Higgs doublet, takes part in charged current interactions. The two-Higgs doublet model (2HDM) \cite{Branco:2011iw} or minimal supersymmetric model (MSSM) are perhaps the simplest models that can have a charged Higgs boson. The recent review volume \cite{Akeroyd:2016ymd} gives a detailed discussion of the charged Higgs boson $H^{\pm}$ in terms of possible models (according to quarks and leptons) and experimental bounds. In all searches, the goal is to disentangle the charged Higgs effects from the background, and this becomes a challenging task for light ($m_{H^{\pm}} \gtrsim 100\ {\rm GeV}$), and weakly-coupled charged Higgs particles.

\begin{figure}[htbp]
\centering
\includegraphics[width=.5\textwidth
]{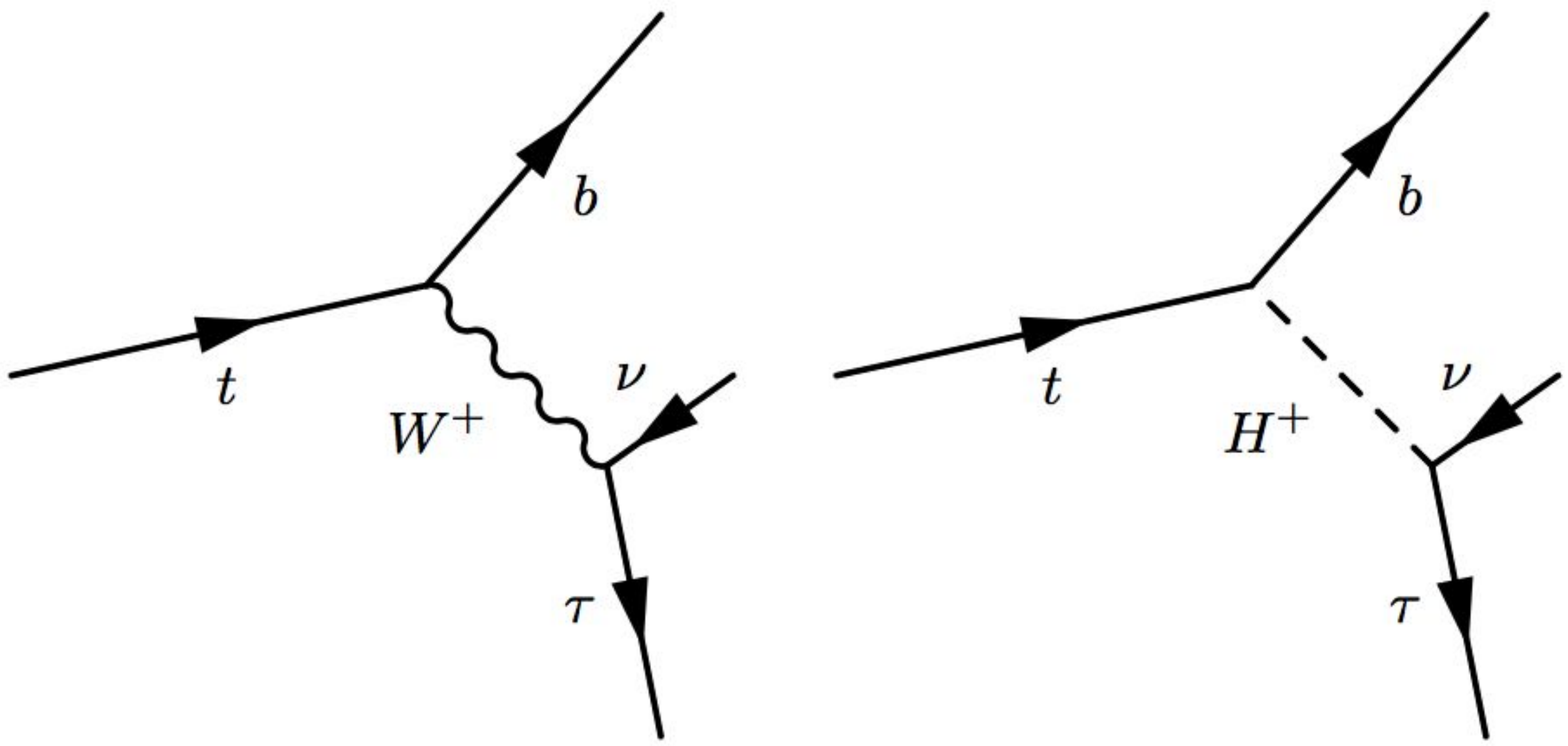}
\caption{\label{fig:hdecay_feynman} Top quark decay through $W^\pm$ and charged Higgs ($H^\pm$) exchanges. The $W^\pm$ diagram generates the irreducible SM background.}
\end{figure}

The golden channel for a light charged Higgs is the top quark decay. This channel is illustrated in Fig. \ref{fig:hdecay_feynman}, where the $W^{\pm}$ (background) and $H^{\pm}$ (signal) contributions are seen to have identical topologies. This channel has been analyzed in detail in \cite{Guedes:2012eu, Guedes:2012yy, Guedes:2013ila} via cut-based signal extraction methods. 

The LHC, which is essentially a top quark factory, is a natural place to search for the charged Higgs via the top quark decay in Fig. \ref{fig:hdecay_feynman}. It is in fact for this reason that method of signal extraction becomes an important issue. Indeed, if the data analysis method is powerful enough to reveal weak signals in huge and mostly-irreducible backgrounds, then it becomes possible to search for elusive particles. The low-mass charged Higgs, whose effects can be shadowed by the W-boson contribution, is difficult to disentangle using cut-based methods. To increase precision, therefore, many machine learning methods that use data to learn, generalize, and predict have been applied. Artificial neural networks (ANN), biology-inspired computational methods, are one of them. However, single-layer ANNs exhibit difficulties in learning particle properties.

To this end, deep neural networks (DNN) \cite{Baldi:2014kfa, Farbin:2016jll, Prosper:2017sla, Schwartzman:2016jqu}, endowed with various layers, showed up as more effective alternatives to single-layer ANNs.  In fact, deep learning, as a new analysis method \cite{Baldi:2014kfa, Farbin:2016jll, Prosper:2017sla, Schwartzman:2016jqu}, started having interesting applications in Higgs physics \cite{Baldi:2014pta, Barberio:2017ngd}, jet physics \cite{Komiske:2016rsd, Erdmann:2017hra, Searcy:2015apa, deOliveira:2015xxd}, string landscape \cite{Gan:2017nyt, He:2017aed}, and astrophysics \cite{Huertas-Company:2015yba, Lanusse:2017vha, George:2017fbn}. Their ability to learn from the raw data, to tackle complex data structures, and to reveal anomalies make DNNs particularly useful. In the present work, we shall apply deep learning to light charged Higgs search with the aim of disentangling it from the irreducible W-boson contribution.

In Sec. II below we will discuss couplings of the charged Higgs boson within 2HDM \cite{Branco:2011iw}. In Sec. III we will discuss data generation and related problems. In Sec. IV we will deep-learn light charged Higgs and contrast our results with those of the cut-based methods \cite{Guedes:2012eu, Guedes:2012yy, Guedes:2013ila}. In Sec. V, we conclude. 

\section{Short Review of 2HDM}

    Extracting a charged Higgs boson information from single top quark decays is the main problem in this study. The simplest model that contains a charged Higgs boson is obtained by adding new Higgs field with the same hypercharge. This model is the simplest extension of the scalar sector of the Standard Model and is it called the 2HDM \cite{Branco:2011iw}. The model contains a total of five Higgs bosons: The neutral ones $h$, $H$, $A$, and the charged one $H^\pm$. These new Higgs states gives rich phenomenology and prominent properties to the model. The 2HDM comes in different types depending on how quarks and leptons interact with each of the Higgs fields. In the Type-X 2HDM \cite{Akeroyd:2016ymd, Branco:2011iw, Guedes:2012eu, Guedes:2012yy, Guedes:2013ila}
    \begin{eqnarray}
    t_R b_L H^{+} :\ -\frac{i \sqrt{2} m_t}{v}\zeta_u\,, \;\; \tau_R \nu_L H^{+} :\ -\frac{ i \sqrt{2} m_{\tau}}{v} \zeta_l
    \end{eqnarray}
    where $\zeta_{u,l}$ is the universal alignment parameters in the flavour space \cite{Abbas:2015cua}, and it is defined as $\zeta_l=-\tan\beta$ and $\zeta_u=\cot\beta$. $\tan\beta$ is the ratio of the vacuum expectation values of the two Higgs doublets. 
    The Type-X 2HDM differs from the others (Type-I, II and Y) by the fact that a different alignment is set in the flavour space and the charged Higgs vertices such as in Fig.\ref{fig:hdecay_feynman} differ. We therefore focus on CP-conserving Type-X 2HDM with a light charged Higgs boson
    \begin{eqnarray} 
    100\ \gev \leq m_{H^{\pm}} \leq  120\ \gev
    \end{eqnarray} 
    as allowed by the current experimental bounds (actually, $m_{H^{\pm}}$ can be appreciably lower; see the recent talk \cite{Moretti:2016qcc}).
    The discovery of the scalar state at the ATLAS and CMS experiments forces us to set one of the neutral states as the Higgs boson. Thus, $h^0$ state is defined to be the SM-like Higgs boson. That is called the exact alignment limit, and the mixing angle among the CP-even Higgs states becomes as $\sin(\beta-\alpha)=1$. Consequently, $h^0$ became indistinguishable from the Standard Model Higgs boson and its mass is set to $m_{h^0}=125\gev$. Besides, $m_{H^0/A^0}=150\gev$ \cite{Akeroyd:2016ymd, Eriksson:2009ws}, the ratio of the vacuum expectation values is set as $\tan\beta= 2$. {It should be kept in mind that small variations in couplings (say, $\tan\beta= 3$, $\sin(\beta-\alpha) = 0.7$) does not have any significant effect on the allowed mass ranges \cite{Guedes:2012eu, Guedes:2012yy, Guedes:2013ila, Moretti:2016qcc}.}
    
    The search for $H^\pm$ has been performed by previous experiments at the LEP through pair production \cite{LEPHiggsWorking:2001aa, Barate:2000qg, Abdallah:2003wd, Acciarri:1999uf, Ackerstaff:1997qk, Abbiendi:2013hk} and at the Tevatron via the decays of the top quarks \cite{Affolder:1999au, Peters:2008wc, Abazov:2008rn, Abazov:2009ae, Abazov:2009wy}.
    The CMS and ATLAS experiments at CERN have been searching it in top quark decays \cite{CMS:2011iqa, Chatrchyan:2012vca, CMS:mxa, Aad:2012tj} in the context of supersymmetry by applying cuts on conventional observables in the detector such as the number of leptons and jets at the final-state as well as kinematical constraints on phase space of reducible and irreducible backgrounds. Light charged Higgs is produced in decays of the top quark. The single top quark cross section associated either with $b$-quark or $W^+$ boson is around $3.04\pb$ at NLO \cite{Aaltonen:2014ura}. Our study involves three broad steps:
    \begin{enumerate}
    \item Production of top quarks via $pp\rightarrow t\; X$ at the LHC energy.
    \item Decay of the top quark as in Fig.\ref{fig:hdecay_feynman}.
    \item Extraction of the charged Higgs information from the data.
    \end{enumerate}
    and this is the program that we will follow below.

\section{Data Generation and Feature Selection} \label{sec:feature}

    We analyze the charged Higgs production and its leptonic decay as a function of its mass (by taking $m_{H^\pm}= 90$, $100$, $110$ and $120\gev$). High-energy proton-proton ($pp$) collisions are simulated at the LHC energies ($\sqrt{s}=13\tev$) using event generators. 
    First, all the couplings and decay widths in the 2HDM are calculated (also as a function of the $\mhp$) for the given parameters defined before. 
    Second, with the help of the relevant files for each $\mhp$ mass value, the proton-proton collisions are simulated using \texttt{MADGRAPH} \cite{Alwall:2011uj}. The production of $t\bar{t}$ in $pp$-collisions at the LHC is the primary source of top-quarks. However, the production of the single-top is an important addition to the charged Higgs production and merits investigation, especially employing the neural network techniques. Then, the top-quark is allowed decaying with charged Higgs and b-quark with $BR(t \rightarrow bH^+)=0.1120$. In Table \ref{tab:tab1} the single-top production process which has the same event topology for the charged Higgs searches are given. Note that, comparison among the production cross sections for the single-top shows that s-channel is negligible. 
    \begin{table}[htp]
      \caption{
    The cross section of the single-top process at $\sqrt{s}=13\tev$ in LHC. \label{tab:tab1} }
    \centering
    \begin{tabular}{l|ccc}
        Process     & s-chan.                        & tW-chan.         & t-chan        \\ \hline
        single top    & 0.549 \cite{Kant:2014oha}        & 71.700 \cite{Kidonakis:2010ux,Kidonakis:2013zqa,Kant:2014oha}        & 216.99 \cite{Kant:2014oha}
    \end{tabular}
    \end{table}
    
    The charged Higgs, characterized by its leptonic decay $H^{+}\rightarrow \overline{\tau} \nu_{\tau}$ is produced in single top decay as in Figure \ref{fig:hdecay_feynman}, where the top quark is produced in $s$-channel, the $tW$ single-top channel, and the $t$-channel in $pp$-collisions. The decay of the charged Higgs is realized in PYTHIA \cite{Sjostrand:2014zea}. Since, $W^{-}$ decays into $e \nu_e$, $\mu \nu_{\mu}$, $\tau \nu_{\tau}$ universally, the single top-quark decaying via $W\pm$ becomes the irreducible background in the Standard Model. Then, we focused on the leptonic branchings of the $\tau$ lepton. Consequently, the charged Higgs signal of interest takes the form 
    \begin{eqnarray}
    p\,p &&\rightarrow  {\text{b-jet}} +   {\text{N-jets}} + {{\ell}} + \slashed{E}
    \end{eqnarray}
    where $\ell = e, \mu$, and $\slashed{E}$ is the missing transverse energy (taken away by neutrinos).

    The relevant background processes, which are considered reducible (at least by applying cuts),  for the charged Higgs searches are given in table \ref{tab:tab2}. These processes are simulated with the help of \texttt{MADGRAPH}, MLM matching algorithm is used to avoid double counting in the production of samples which includes additional jets. Besides, the background samples are produced with the following features; all jets (including b-jets) and leptons have transverse momentum $p_T>20 \gev$, pseudorapidity $|\eta|\leq 2.4$, radial angle $\Delta R>0.4$, and missing transverse energy due to the neutrinos in each event $\text{MET}>20\gev$.
    \begin{table}[htp]
    \caption{
    The collection of the processes and their cross sections in the Standard Model which contributes to the charged Higgs searches at $\sqrt{s}=13\tev$ in LHC. The cross section values are given in pb. \label{tab:tab2}}
    \centering
    \begin{tabular}{l|ccc}
    The process			& +0 jets	& +1 jets		& +2 jets	\\ \hline
    $W^\pm$			& 24390		& 4288			& 1637		\\
    $W^\pm c$		& 586.0		& 503.0			& 269.6		\\
    $W^\pm bb$		& 5.01		& 8.5			& 9.6		\\ \hline\hline
					& leptonic	& semi-leptonic	& hadronic	\\ \hline
    $t\bar{t}$		& 22.45		& 69.78			& 116.16	\\ 
    \end{tabular}
    \end{table}
    Naturally, the irreducible background would be coming from the diagram on the left in figure \ref{fig:hdecay_feynman} (the W-boson contribution) is also produced. All the events are dressed by simulating the fragmentation, parton shower and hadronization stages on quarks to form jets. This step is performed using \texttt{PYTHIA} \cite{Sjostrand:2014zea}. Finally, detector simulation is performed using the fast simulation software \texttt{Delphes-3.4.1} \cite{deFavereau:2013fsa}, where detector and trigger configurations are set to the CMS defaults.

    In the course of reconstruction, all objects (jets, leptons and missing transverse energy) are defined through their kinematical variables, that is, the transverse momentum ($p_T$), rapidity ($\eta$) and azimuthal angle ($\phi$). These kinematic variables are registered for all objects in each event. Following the reconstruction, it is necessary to set the features characterizing the targeted event, that is, the charged Higgs identification. The essential feature is that an event must have one and only one isolated lepton (coming from $\tau$ decays) if it is to have anything to do with charged Higgs signal. Besides this, $b$-tagging can be included to the extent that its uncertainty is not worse than that in the DNN search. This preselection stage eliminates part of the events registered. In general, efficiency is measured by the signal-to-background ratio ($S/B$).

    In the present work, the performance of the DNN structure on the charged Higgs identification problem will be investigated by using two different features sets. The Feature Set 1 (FS1) mostly involves kinematic variables pertaining to a single particle or highest-energy jet. 
    It is composed of the followings:

\begin{itemize}
\item {\it Feature Set 1: } 
\begin{enumerate}
\item $p_T^{\ell}:\ $ transverse momentum of the lepton (muon or electron). 
\item $\slashed{E}_T:\ $ Missing Transverse Energy (MET) of neutrinos,
\item $N^{jets}:\ $  number of jets in an event,
\item $N^{b-tag}:\ $  number of $b$-tagged jets in an event,
\item $\Delta R_{\ell b}:\ $  distance between the lepton and the $b$-tagged jet (distance or angle),
\item $\Delta\eta_{\ell b}:\ $  relative rapidity between the lepton and $b$-tagged jet (distance or angle),
\item $p_T^{jet-1}:\ $ transverse momentum of the highest-energy jet,
\item $\eta^{jet-i}:\ $  rapidity of the highest-energy jet,
\item $\phi^{jet-i}:\ $  azimuthal angle of the highest-energy jet,
\item $b$-tag:  $b$-tagging information,
\item $\eta_{\ell}:\ $ lepton rapidity,
\item $\phi_{\ell}:\ $ lepton azimuthal angle,
\item $\eta^{\slashed{E}_T}:\ $ MET rapidity,
\item $\phi^{\slashed{E}_T}:\ $ MET azimuthal angle,
\item $p_T^{b-tag}:\ $ transverse momentum of b-tagged jet.
\end{enumerate}

The Feature Set 2 (FS2) mostly involves combinations of different features in FS1. These features, in general, are based on particle physics relations pertaining to charged Higgs production and therefore are called human engineered features. It is given by 

\item {\it Feature Set 2} 
\begin{enumerate}
\item $M_T^{\ell\nu}:\ $ transverse mass of leptons (it can be written as  
$M_T^{\ell\nu}=\sqrt{2p_{\ell T}\slashed{p_T} - 2(p_{\ell x}\slashed{p}_x+p_{\ell y}\slashed{p}_y) }$ where $p_{\ell T}$ is the lepton transverse momentum).
\item $M_{\ell\nu b}:\ $ the reconstructed top quark mass,
\item $M_{jj}:\ $  dijet invariant mass (the highest-energy two jets),
\item $M_{jjj}:\ $ trijet invariant mass, 
\item $\alpha_{T}:\ $ parameter controlling the QCD background ($\alpha_{T} = p_T^{jet-2}/m_{jj}$).
\end{enumerate}
\end{itemize}

\begin{figure}
\centering
  \centering
  \subfloat[] {\includegraphics[ width=.45\linewidth ]{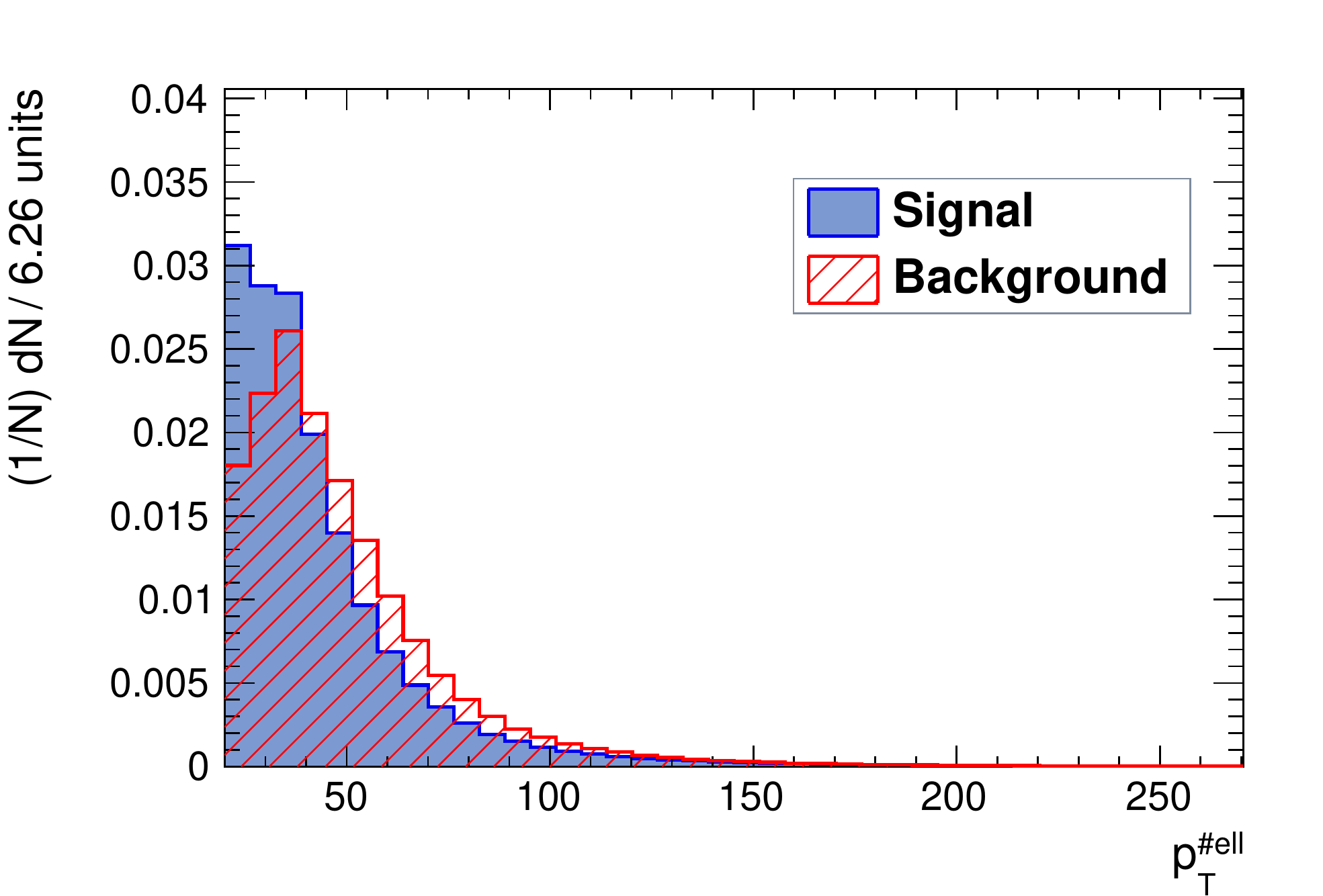}}
   \subfloat[] {\includegraphics[width=.45\linewidth]{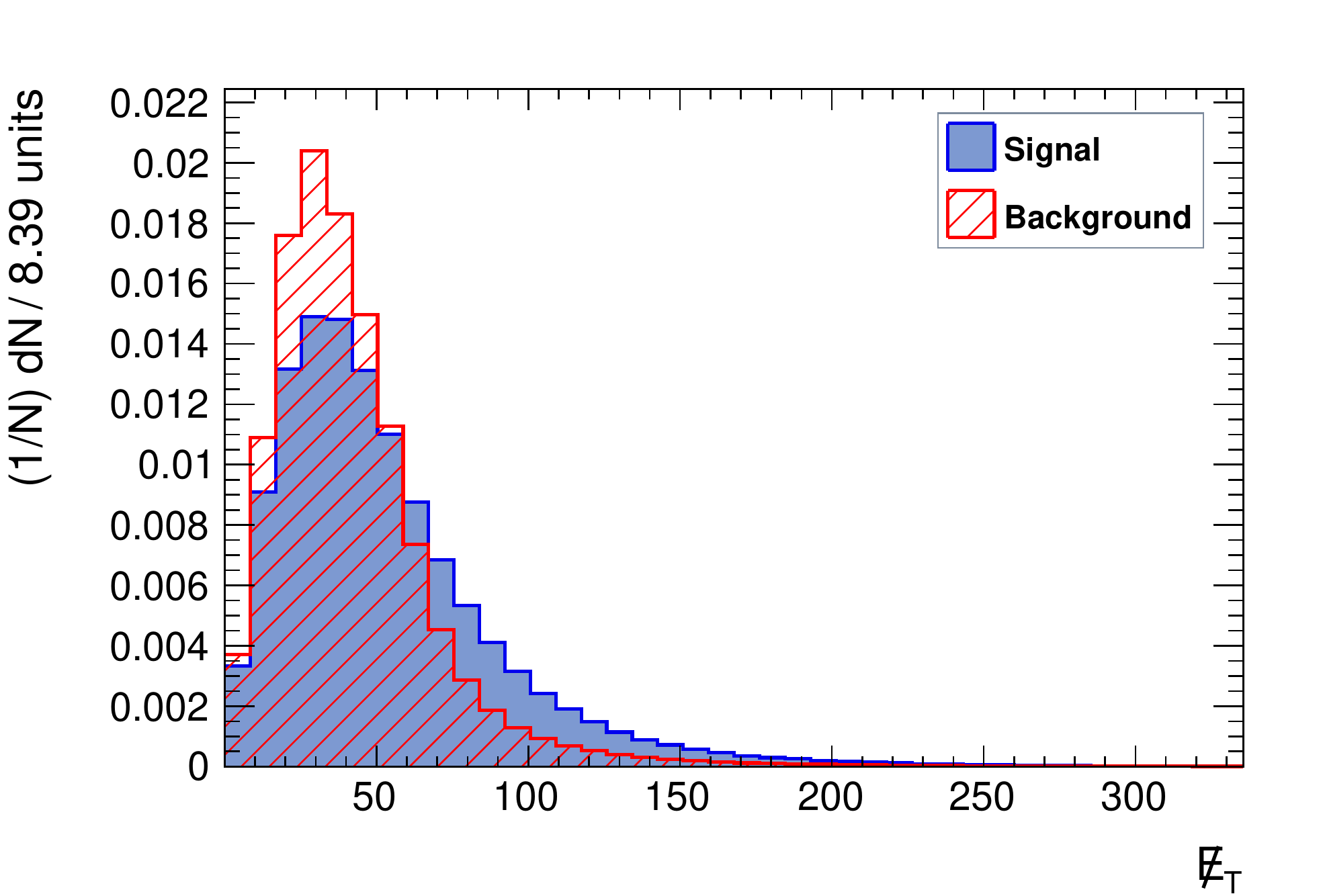}} \\
    \subfloat[] {\includegraphics[width=0.45\linewidth]{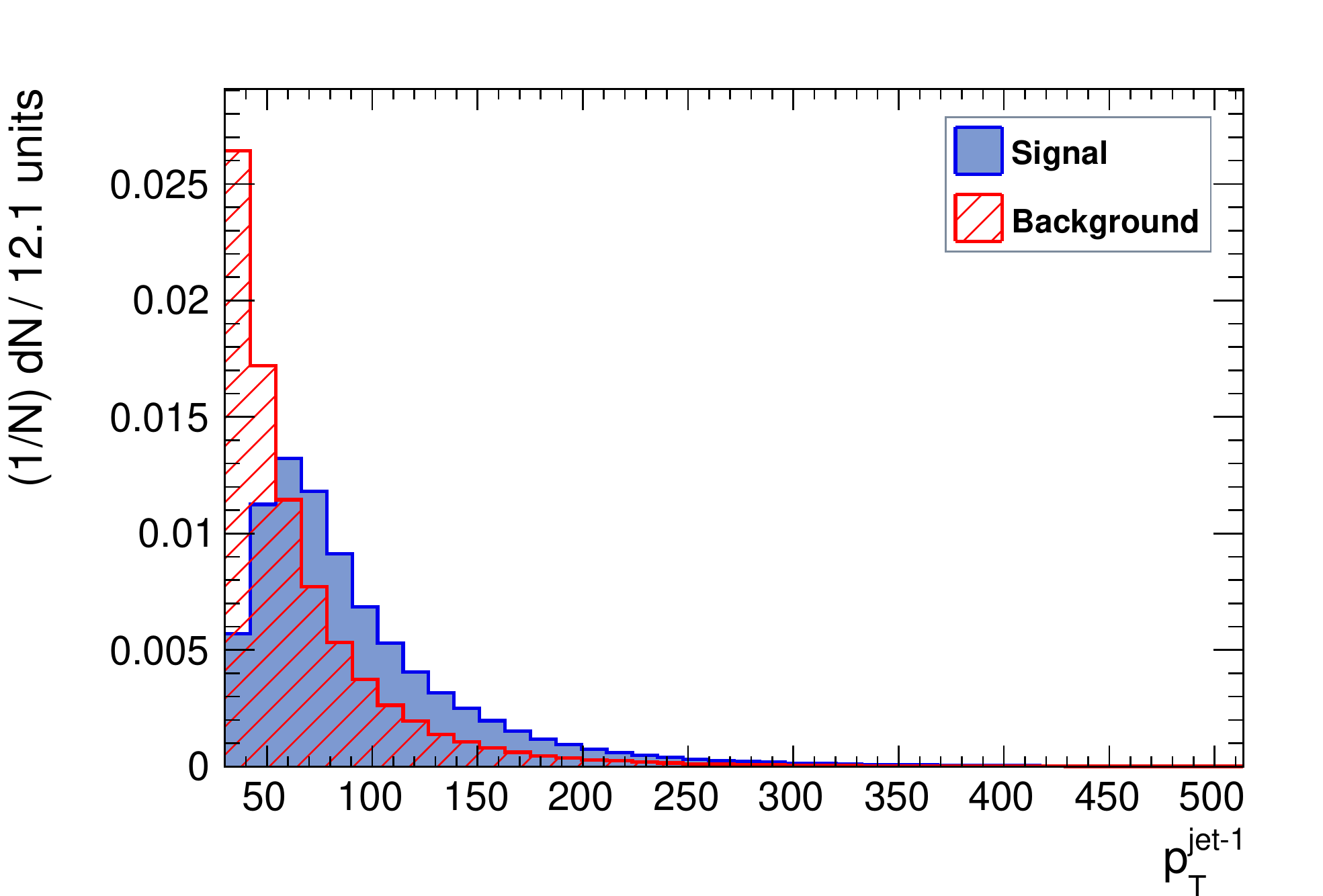}} 
   \subfloat[] {\includegraphics[width=.45\linewidth]{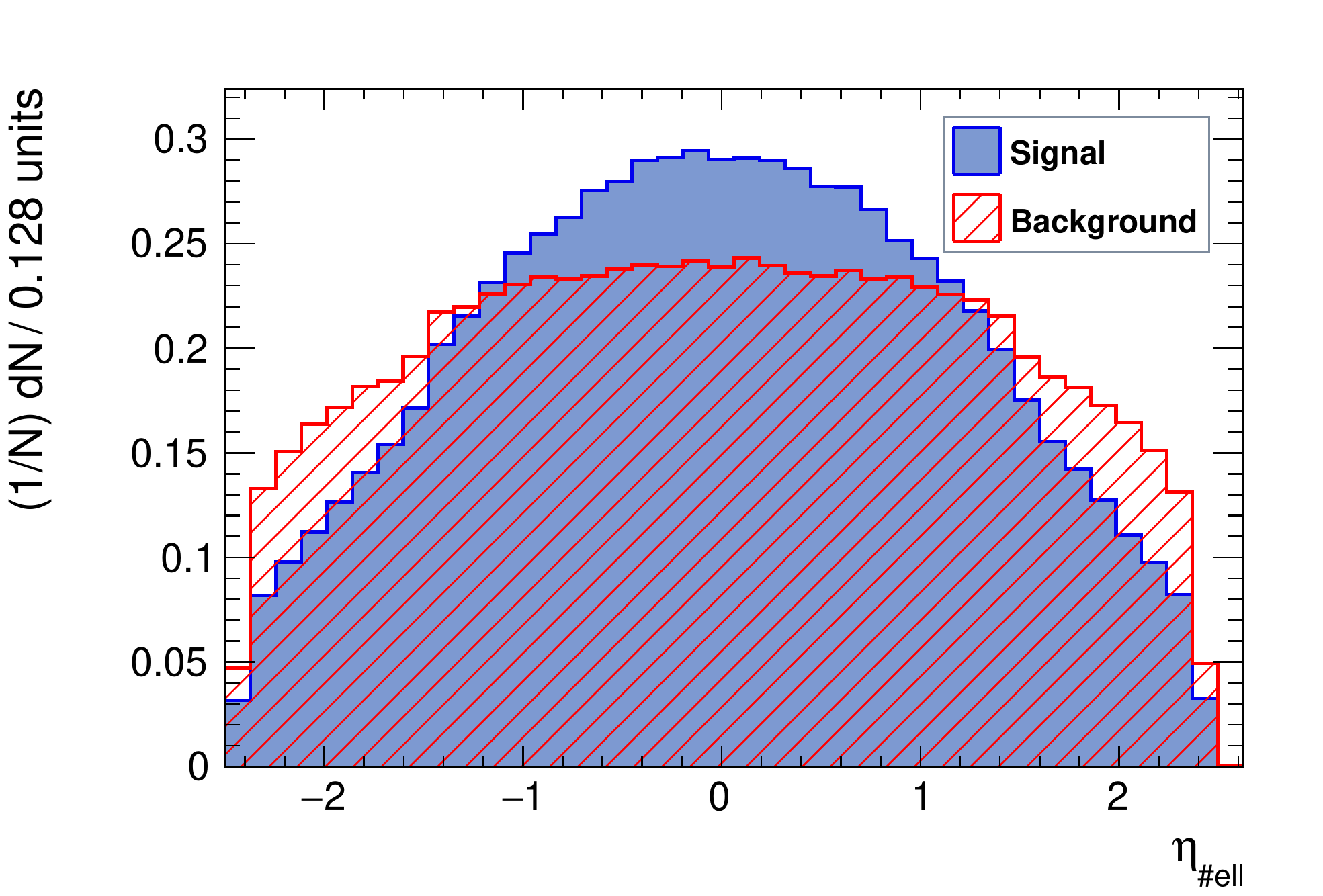}} \\
    \subfloat[] {\includegraphics[width=0.45\linewidth]{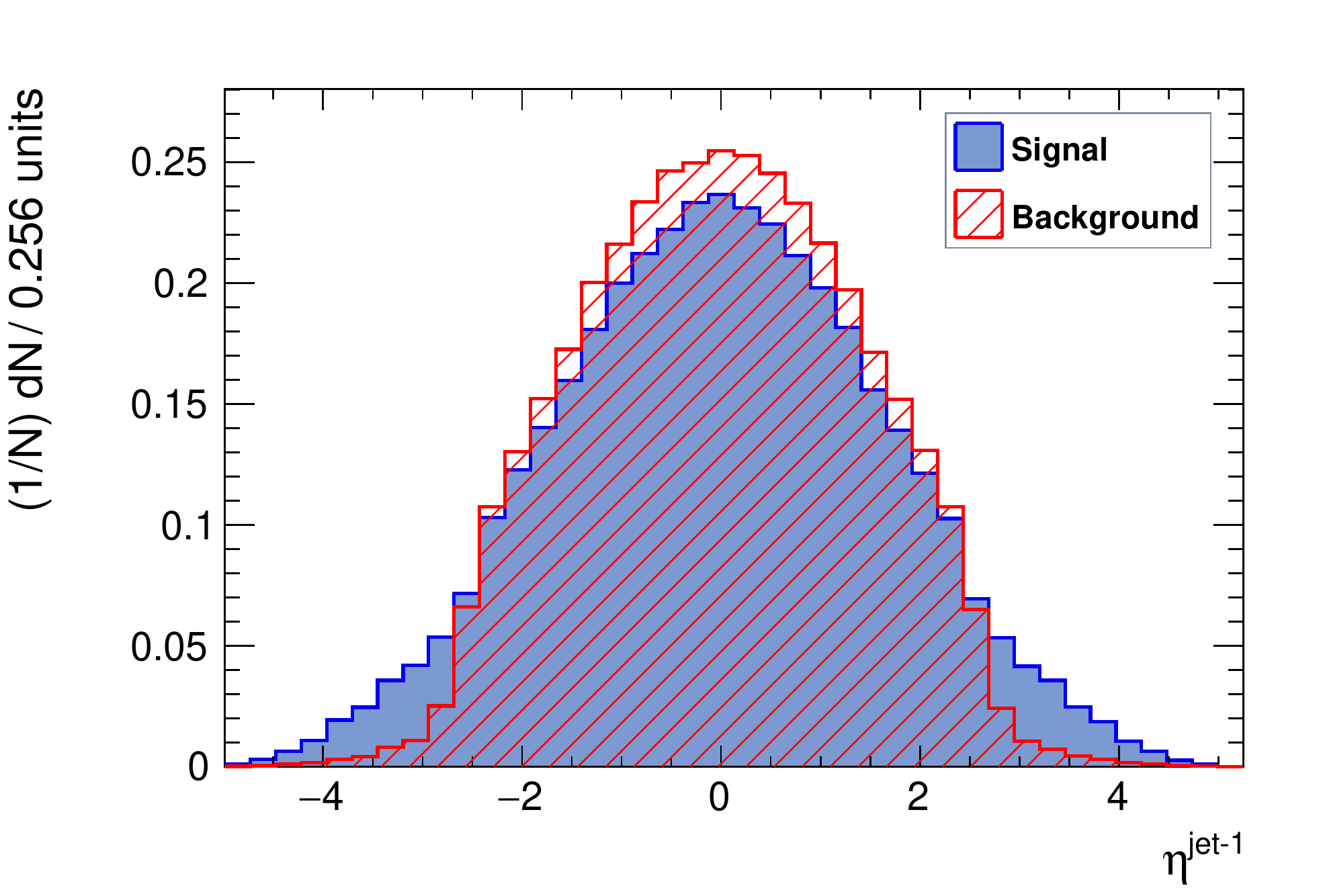}}
   \subfloat[] {\includegraphics[width=.45\linewidth]{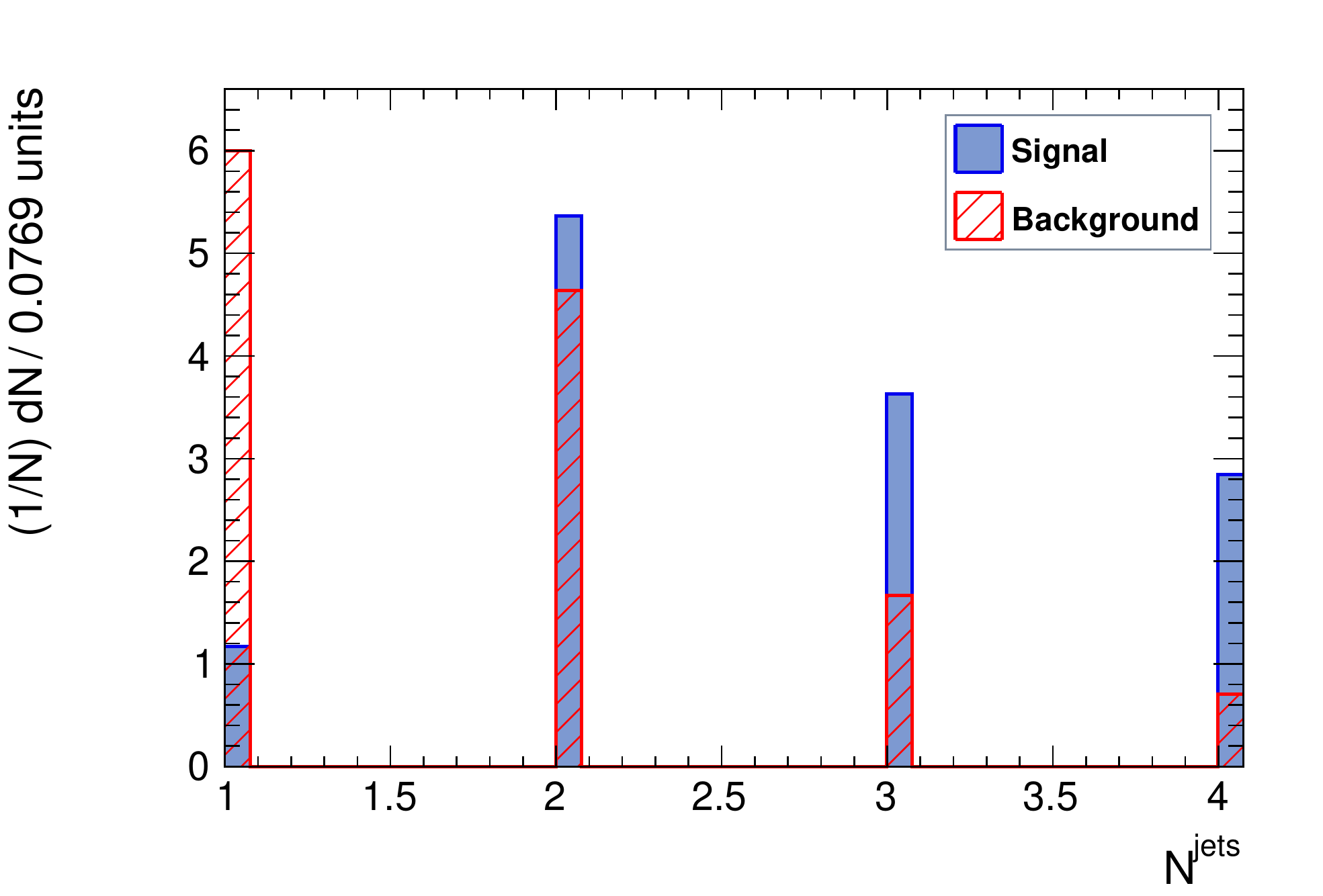}}
  \caption{Distributions of some features ($p^{\ell}_{T}$, $\slashed{E}_T$, $p^{jet-1}_{T}$, $\eta_{\ell}$, $\eta^{jet-1}$, $N^{jets}$) from FS1. The plots suggest that, generally, signal remains buried in the background for all six features (maybe excepting the lepton $\eta$ distribution). Applying cuts on individual features, therefore, may not always lead to an effective signal extraction.}
\label{fig:fs1}
\end{figure}

\begin{figure}
\centering
  \centering
  \subfloat[] {\includegraphics[ width=.45\linewidth ]{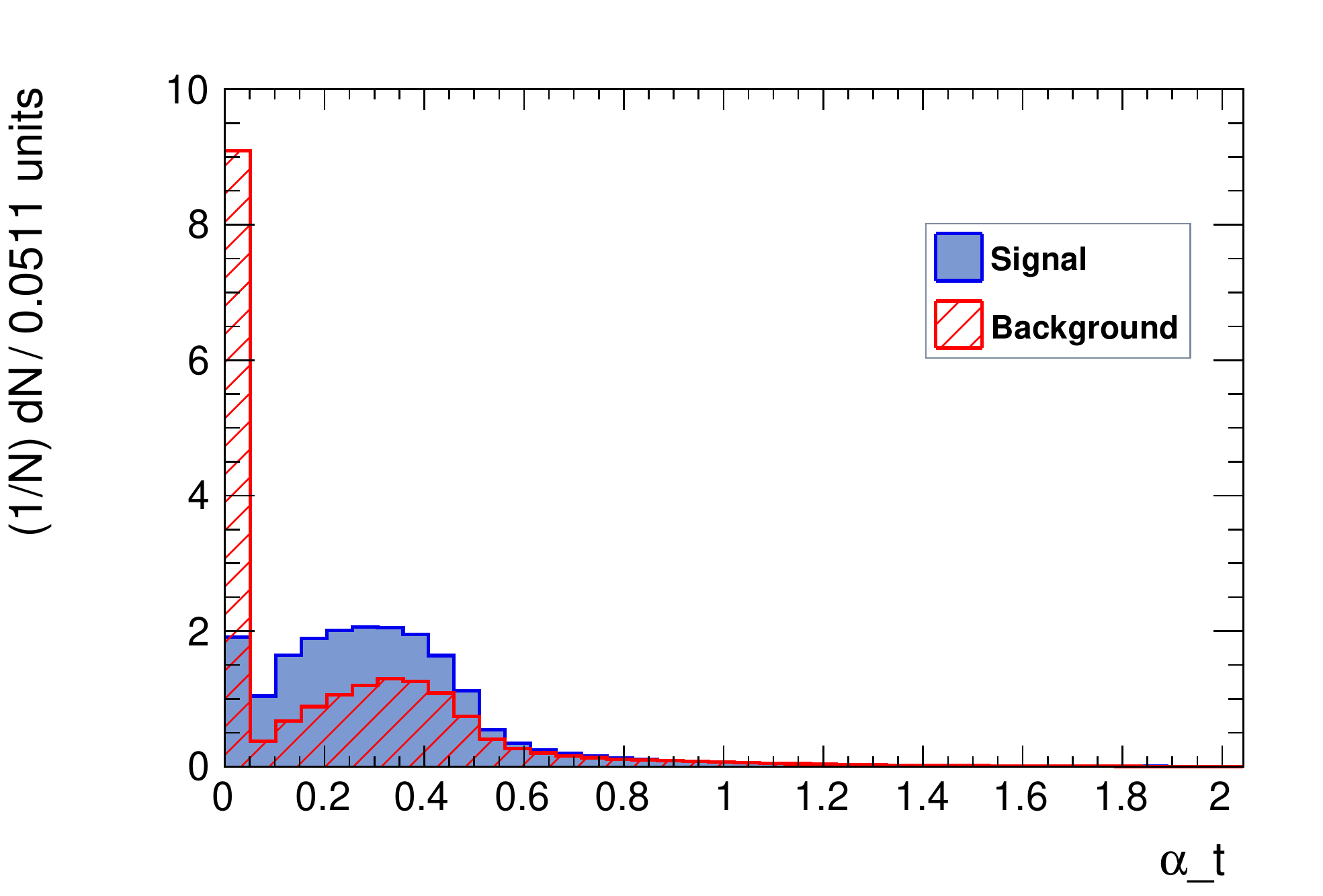}}
   \subfloat[] {\includegraphics[width=.45\linewidth]{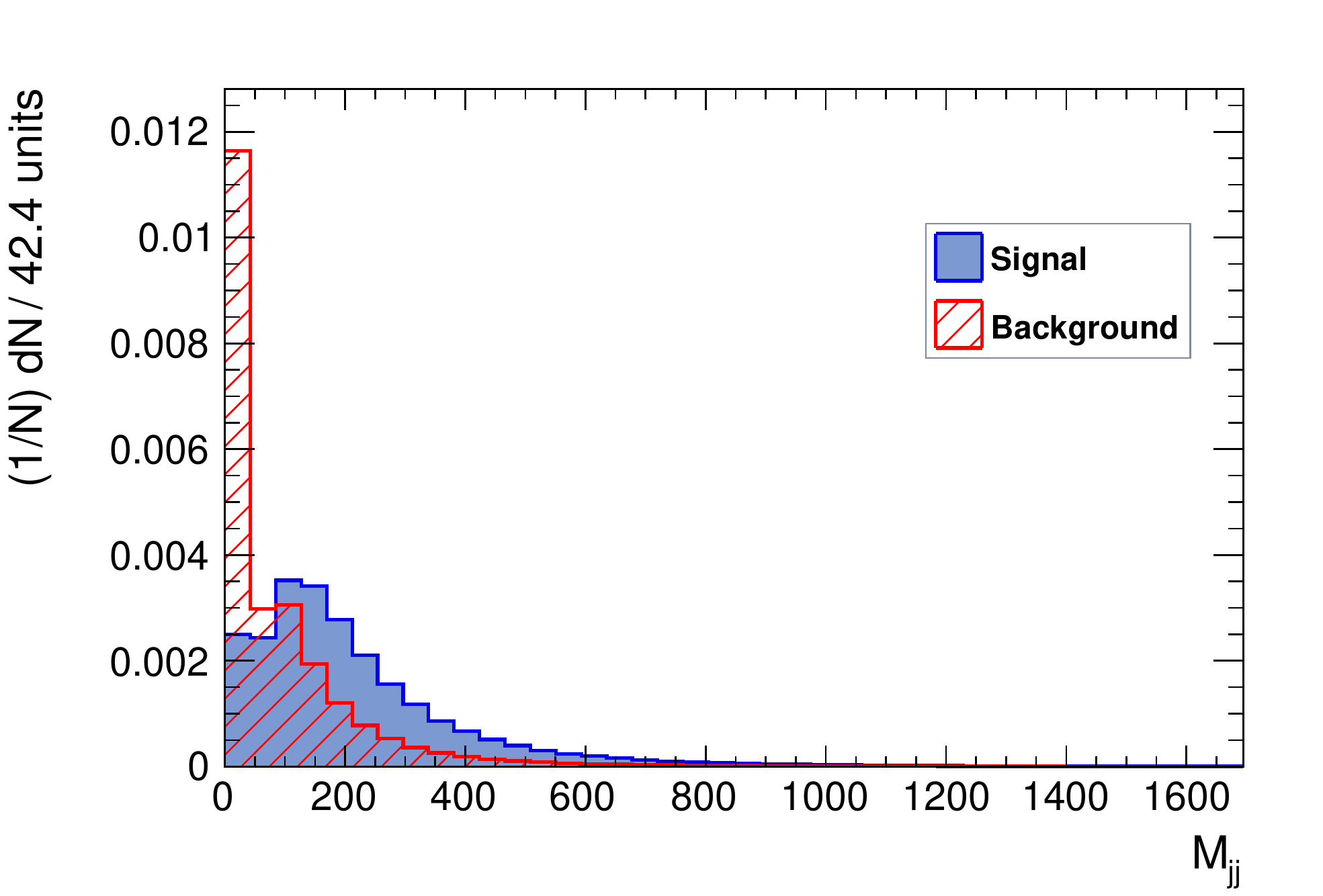}} \\
    \subfloat[] {\includegraphics[width=.45\linewidth]{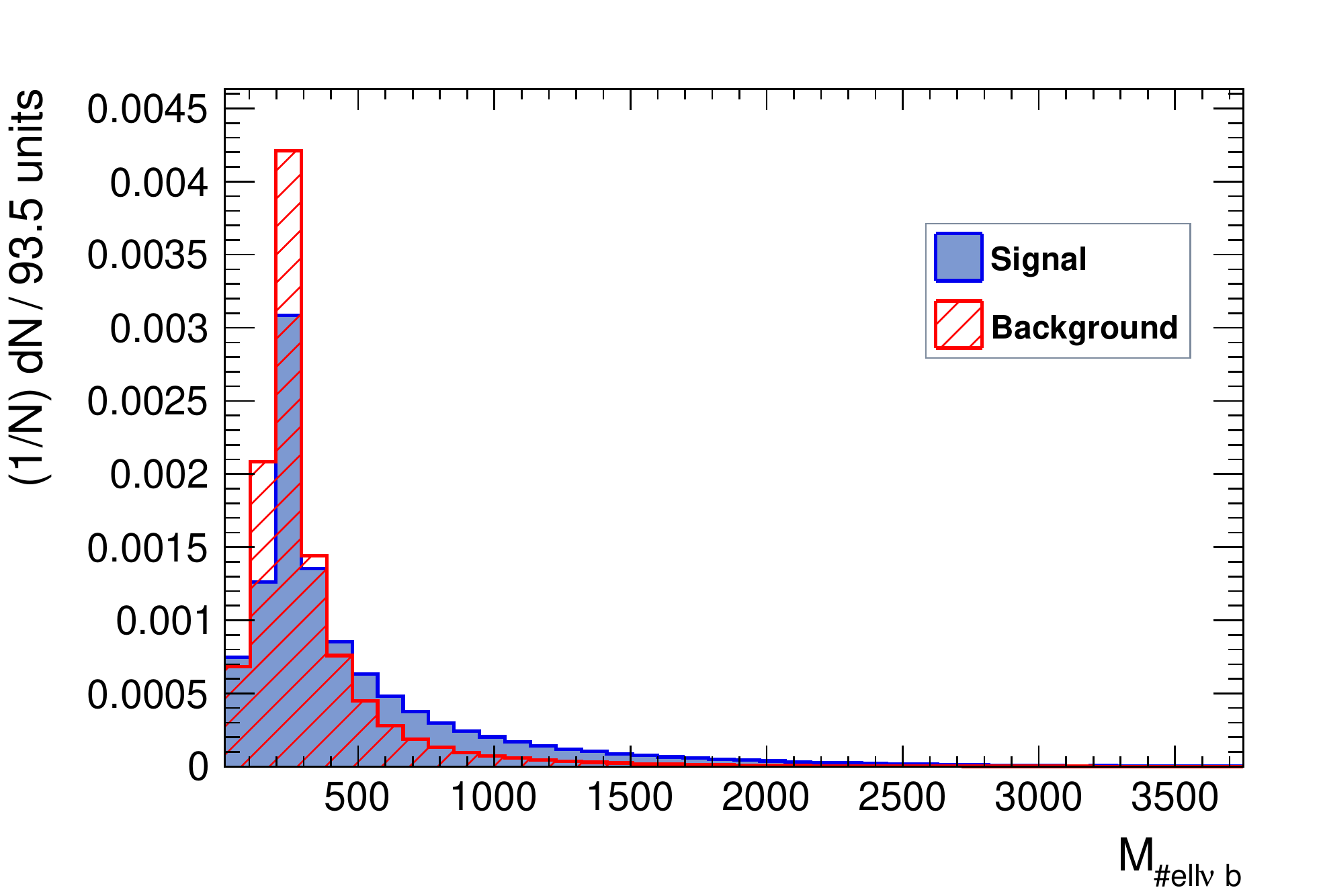}}
    \subfloat[] {\includegraphics[width=0.45\linewidth]{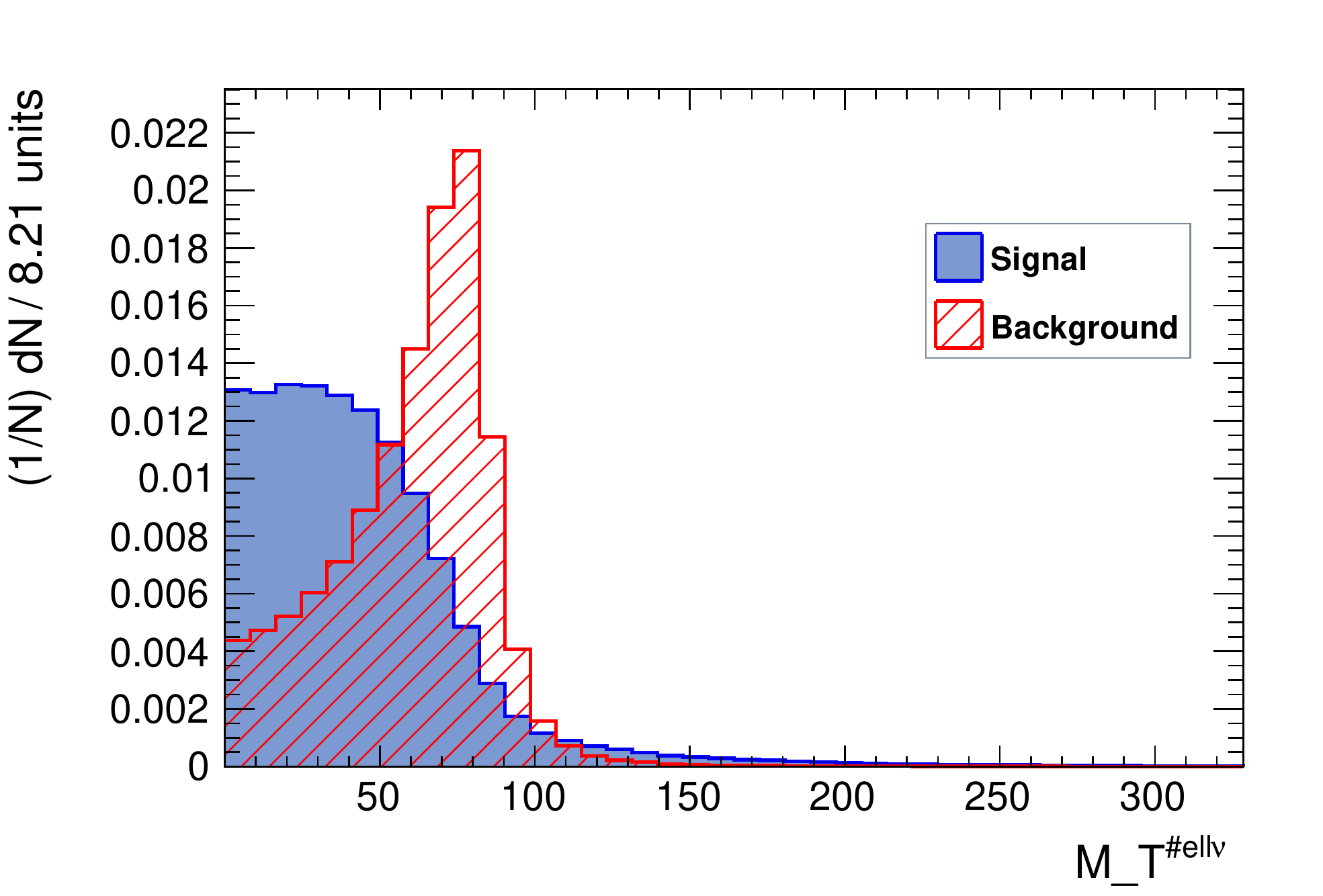}}
     \caption{Distributions of some features ($\alpha_{T}$, $M_{jj}$, $M_{\ell\nu b}$, $M_T^{\ell\nu}$) from FS2. The plots suggest that, generally, signal and background are better separated compared to the FS1 distributions in Figure \ref{fig:fs1}. These features are therefore more eligible for applying cuts.}
\label{fig:fs2}
\end{figure}

It is clear that majority of the features in FS2, for instance, $M_T^{\ell\nu}$, can be expressed as combinations of the features in FS1. Our goal in this work is to determine whether a DNN of appropriate architecture can inherently form high-level discriminative features like those in FS2 as part of its learning process. For this purpose, we will train DNN for both FS1 and FS1+FS2 input features, and contrast their outputs to determine signal efficiencies. To this end, we plot distributions of features for both background and signal. Figure \ref{fig:fs1} shows distributions of some features from FS1. Figure \ref{fig:fs2} does the same thing for features from FS2. In each plot, the background is red-shaded, and the charged Higgs signal is blue-shaded. As follows from the figures \ref{fig:fs1} and \ref{fig:fs2},  the features in FS2 exhibit better discriminative power, compared to those in FS1, in classifying the background and the signal. 
\section{Extracting Charged Higgs with DNN}

DNNs are typically structured as feedforward neural networks. Feedforward DNN architecture consists of multiple layers of processing units the input and the output. It has the capability of modeling complex non-linear relationships between the event variables and the signal. It generates hierarchically arranged submodels between the layers as a result of which output is formed as a layered arrangement of the features underlying the signal and background events. Each layer in the DNN allows for arrangement of the features (outputs) arising from the preceding layers.

In this study, DNNs are trained and validated using the generated data set of 2 million samples (Section 3). Classification performance of deep neural networks is evaluated according to its response to different input feature sets and its capability of extracting the underlying characteristics. To this end, we use two performance metrics: {\it(i)} Signal efficiency, and {\it(ii)} $S/\sqrt{S+B}$ where $S$ and $B$ denote the signal and background. 

Considered in this work is two different DNN structures. Each DNN consists of three hidden layers of $N$ hyperbolic tangent units and a linear output unit with cross-entropy loss. The first DNN consists of $N=100$
processing units in each layer and we call it $DNN^{100}$.  The second DNN will be called $DNN^{300}$ because it contains $N=300$ processing units. Xavier weight initialization is used to make sure that the weights lie in acceptable ranges \cite{pmlr-v9-glorot10a}. After grid search, the learning rate and momentum are set at $10^{-5}$ and $0.5$, respectively.  In addition, we add a dropout with a fraction of $0.5$ for the second and third layers. Weight decay is $10^{-5}$. All weights are regularized using L2 regularization multiplied by a weighting factor of $10^{-5}$ to prevent overfitting.  All input features are normalized to zero mean and unit variance. 

 \begin{figure}
\centering
  \centering
  \subfloat[] {\includegraphics[ width=.45\linewidth ]{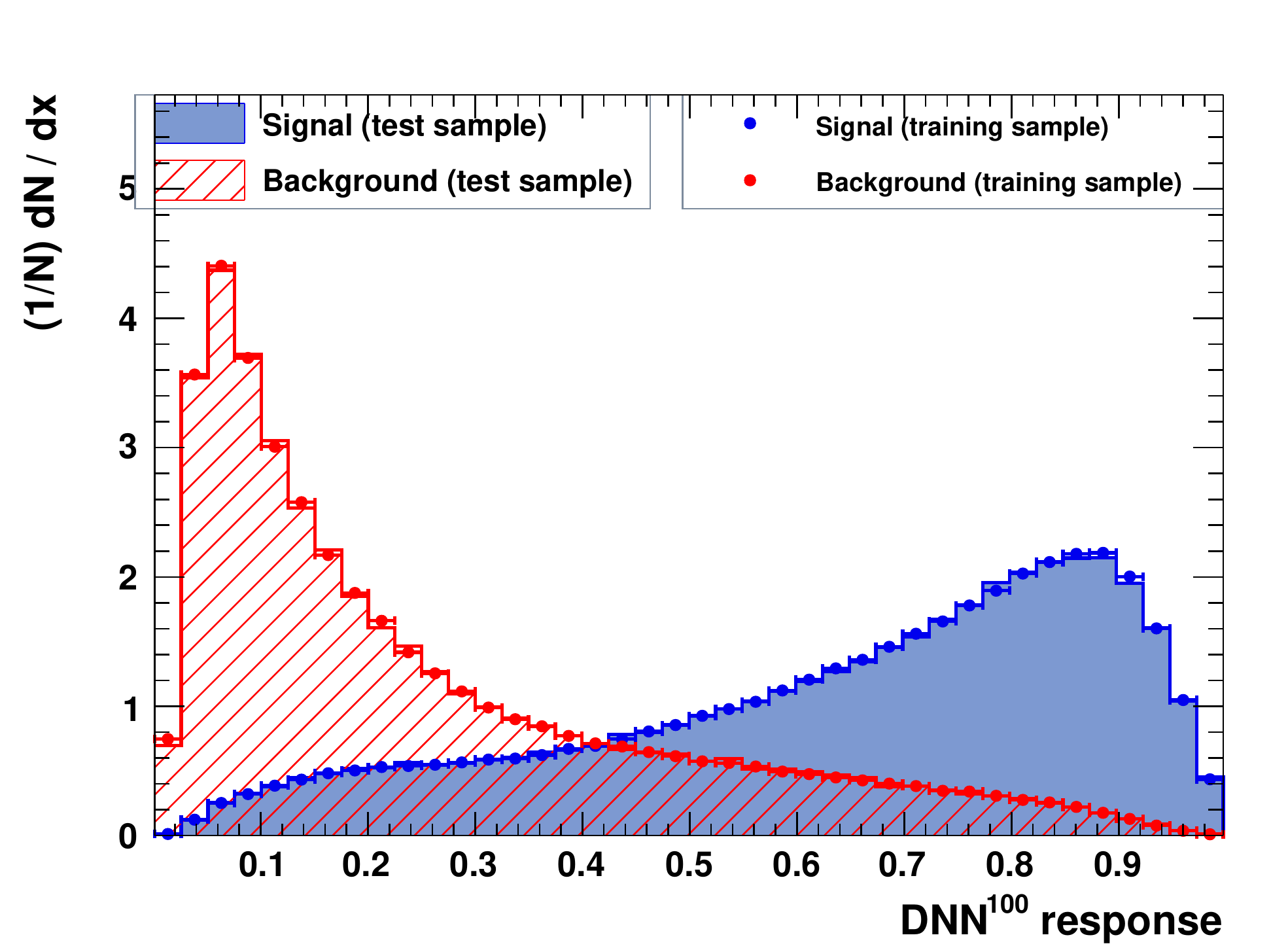}}
   \subfloat[] {\includegraphics[width=.45\linewidth]{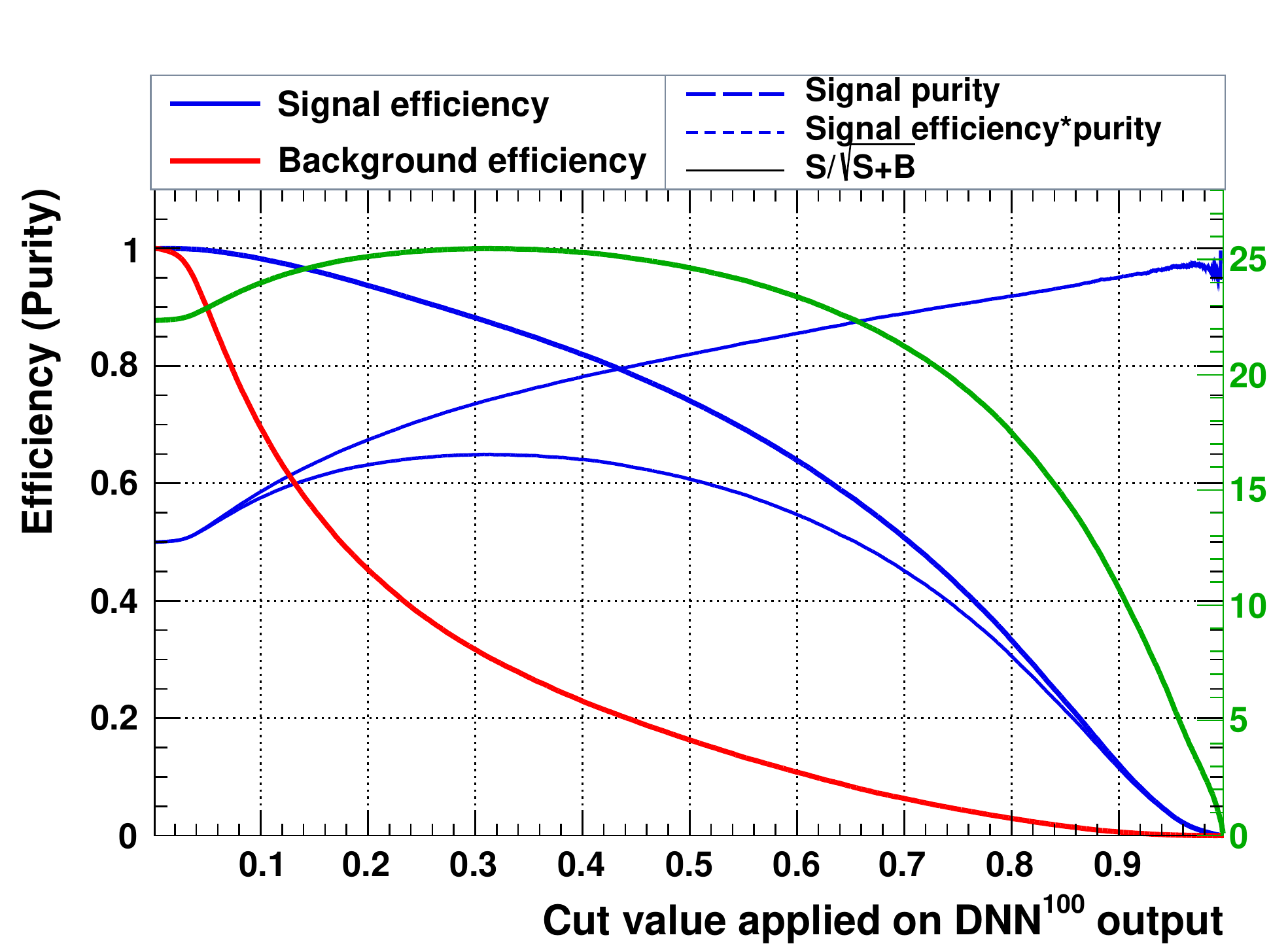}} 
   \caption{(a) Output distributions and (b) cut efficiencies of FS1+FS2 input features for the $DNN^{100}$ classifier with a charged Higgs mass of $m_{H^{\pm}} =100\, {\rm GeV}$.}
\label{output100}
\end{figure}

\begin{figure}
\centering
  \centering
   \subfloat[] {\includegraphics[ width=.45\linewidth ]{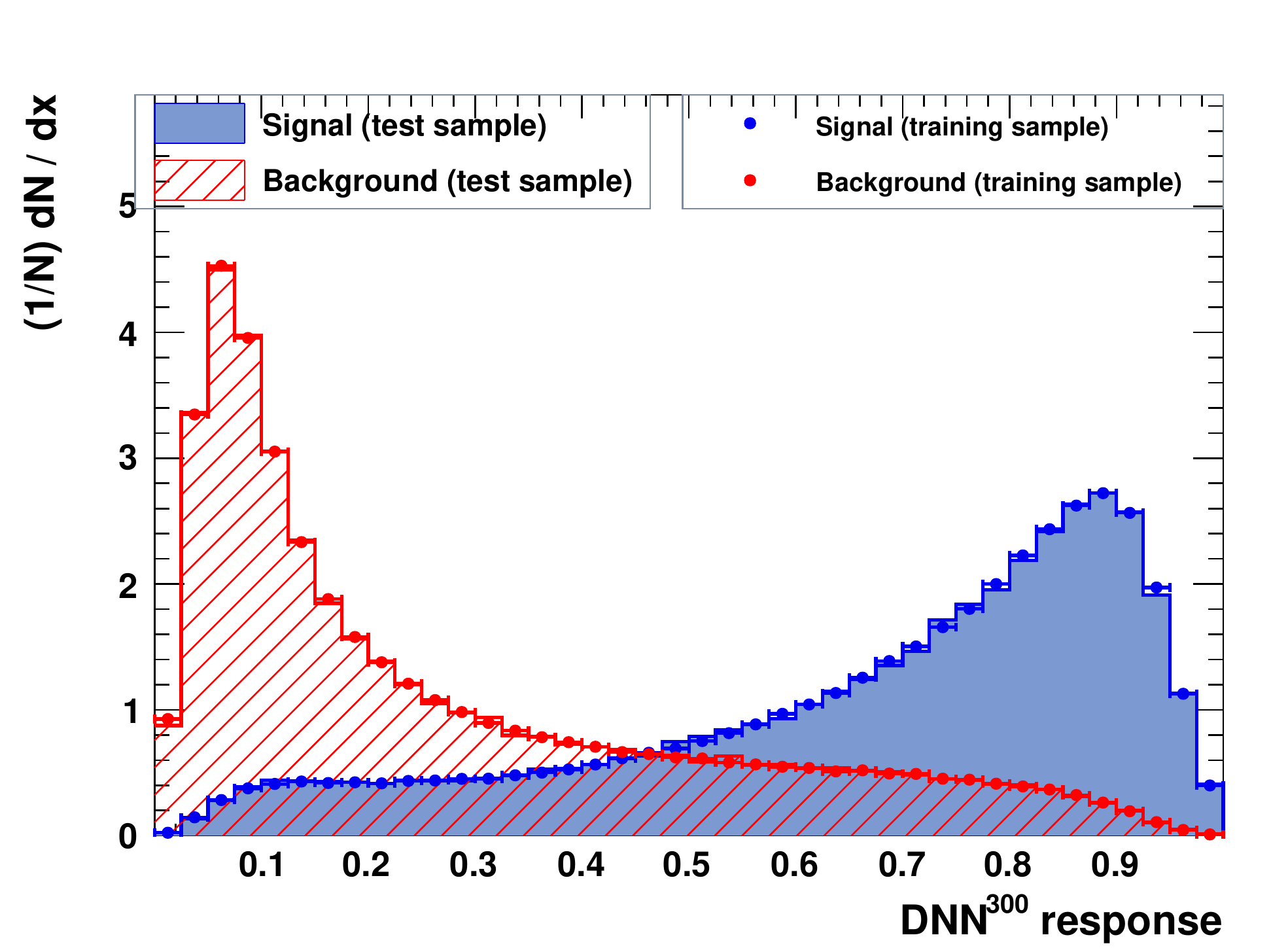}} 
   \subfloat[] {\includegraphics[width=.45\linewidth]{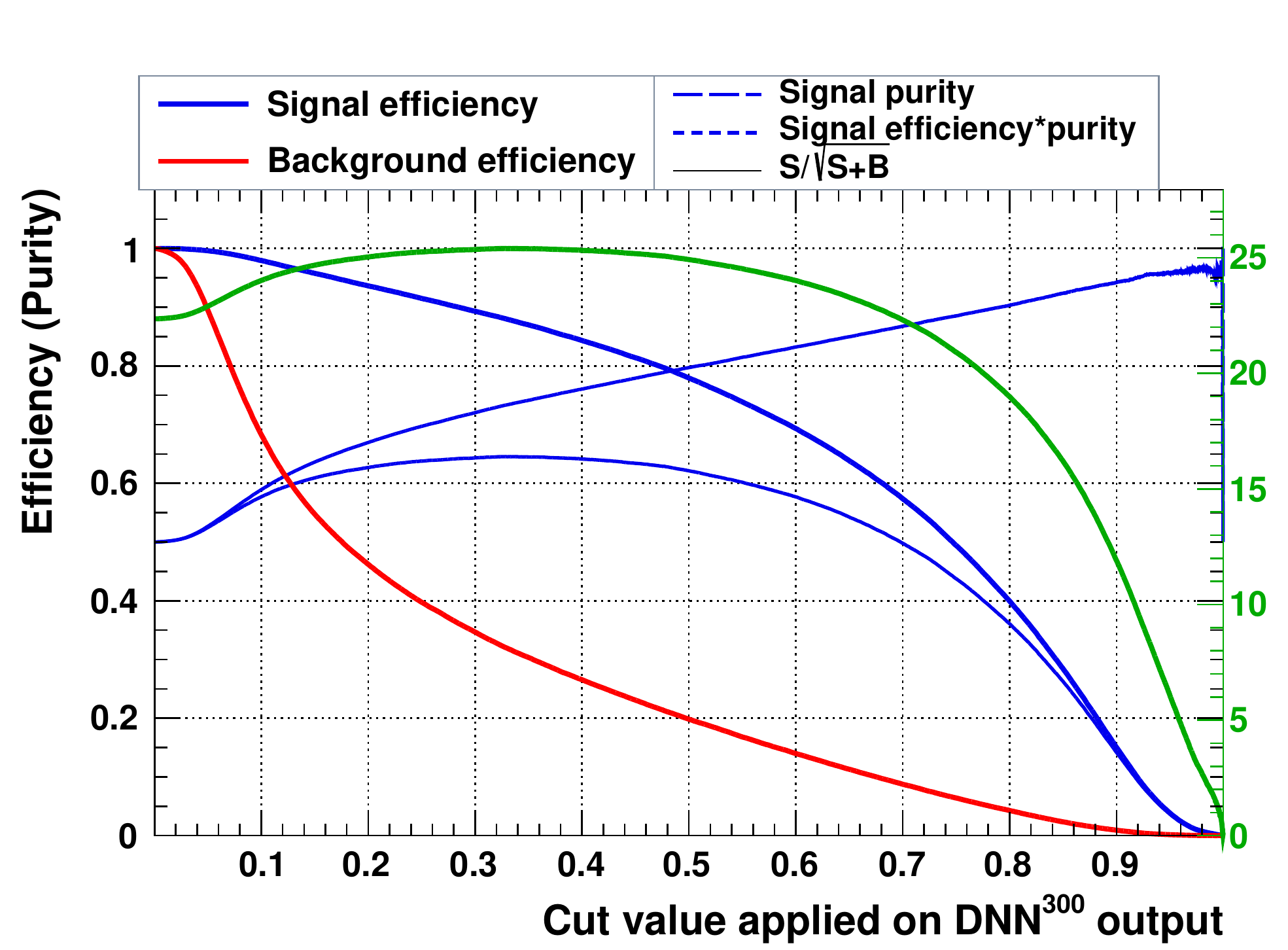}}
     \caption{ (a) Output distributions and (b) cut efficiencies of FS1+FS2 input features for the $DNN^{300}$ classifier with a charged Higgs mass of $m_{H^{\pm}} = 100\, {\rm GeV}$.}
\label{output300}
\end{figure}

In our analysis we explore effects of different charged Higgs masses ($m_{H}=90, 100, 110,$ $120\, {GeV}$), different DNN architectures ($DNN^{100}$ and $DNN^{300}$), different feature sets (FS1 and FS1+FS2), and single-layered neutral network structures (with processing units $N=20,100$ and 300).  

In Figure \ref{output100}, we plot DNN response (a) and cut efficiencies (b). We take FS1+FS2 input features, use the $DNN^{100}$ classifier, and consider a charged Higgs mass of $m_{H^{\pm}} =100\, {\rm GeV}$. It is clear from Figure \ref{output100} (a)  that the signal (blue) and background (red) are well separated. Optimal extraction is achieved for a $DNN^{100}$ response around $0.4$. Depicted in Figure \ref{output100} (b) are the signal and background efficiencies, signal purity, signal efficiency times purity, and $S/\sqrt{S+B}$ as functions of the cut value applied to the $DNN^{100}$ response. Indeed, the signal efficiency is seen to be maximal for a cut value around 0.4. 
 
In Figure \ref{output300}, we repeat the analysis in Figure \ref{output100} by taking $N=300$. The results are similar except for slight decreases in efficiency values and shift of the optimal cut value towards 0.5.

\begin{figure}
\centering
  \centering
  \subfloat[] {\includegraphics[ width=.42\linewidth ]{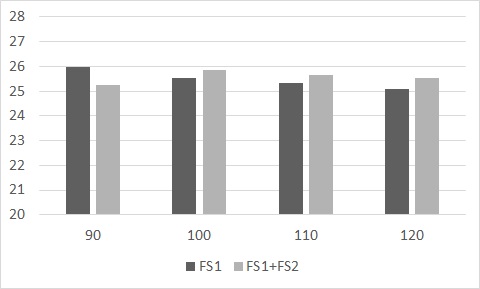}}
   \subfloat[] {\includegraphics[width=.45\linewidth]{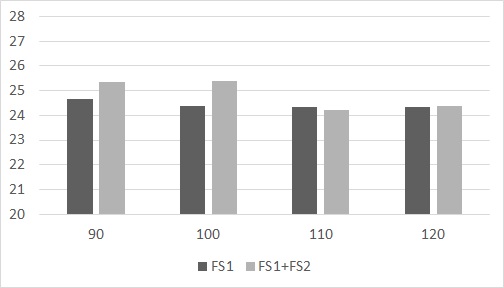}}
     \caption{The metric $S/\sqrt{S+B}$ on FS1 (black) and FS1+FS2 (grey) as a function of the charged Higgs mass $m_H$ for $DNN^{100}$ (a) and $DNN^{300}$ (b) architectures.}
\label{snratio}
\end{figure}

In Figure \ref{snratio}, we depict the metric $S/\sqrt{S+B}$ for each Higgs mass value by considering $DNN^{100}$ (a) and $DNN^{300}$ (b) architectures on FS1 (black) and FS1+FS2 (grey) input features. From this, one arrives at two important observations:
\begin{enumerate}
\item It is noticed that, generally, $S/\sqrt{S+B}$, the signal-to-noise ratio, is higher in $DNN^{100}$ than in $DNN^{300}$ independent of the input feature set. From this, it is seen that increasing the number of processing units in DNN does not proportionally increase efficiency {\it i.e.} $S/\sqrt{S+B}$. This is mainly due to the increased complexity in higher-dimensional learning space. 

\item It is also seen that inclusion of high-level features (switching from FS1 to FS1+FS2) does not lead to any dramatic increase in  $S/\sqrt{S+B}$. This means that DNN (especially $DNN^{100}$) is efficient enough to inherently learn combinations of the input features which increase the efficiency. In essence, the human engineered features in FS2 such as $M_{\ell\nu b}$ (constructed from particle physics perspective) are replaced by the learned non-linear relationships among layers of the $DNN^{100}$. 
\end{enumerate}

The left blocks of Tables \ref{snfs1} and \ref{snfs1fs2} show the signal efficiencies for the optimal cut values used in obtaining $S/\sqrt{S+B}$ in Figure \ref{snratio}. First, parallel to conclusions arrived after Figure \ref{snratio}, signal efficiencies for $DNN^{100}$ are mostly higher than those for $DNN^{300}$. Second, for some Higgs masses, efficiencies are few percent higher in FS1+FS2 than in FS1 alone. Third, $DNN^{100}$ performs better than $SNN^{20,100,300}$ (excepting $m_H=90\ {\rm GeV}$), and its performance increases with increasing $m_H$.

\begingroup
\setlength{\tabcolsep}{6pt} 
\renewcommand{\arraystretch}{1.2} 
\begin{table}[h]
\caption{The signal efficiencies for $DNN^{100}$ and $DNN^{300}$ (left) and shallow neural network (SNN) for
$SNN^{20}$, $SNN^{100}$ and $SNN^{300}$ (right). The input features are FS1.} \label{snfs1}
\centering 
\begin{tabular}{c|rr|rrr} 
\hline\hline 
 $m_{H^{\pm}}$  & $DNN^{100}$& $DNN^{300}$ & $SNN^{20}$& $SNN^{100}$& $SNN^{300}$ \\
\hline
90  & 0.880  &0.8588 & 0.886 & 0.890 &  0.887 \\ 
100 & 0.878   & 0.8741 &0.869 & 0.878 &  0.873   \\ 
110 & 0.874  & 0.8778 &  0.844 & 0.847  & 0.842 \\
120 & 0.897  & 0.8717 & 0.838 & 0.835 &   0.843 \\[1ex] 
\hline 
\end{tabular}
\label{t1}
\end{table}
\endgroup

\begingroup
\setlength{\tabcolsep}{6pt} 
\renewcommand{\arraystretch}{1.2} 
\begin{table}[h]
\caption{The signal efficiencies for $DNN^{100}$ and $DNN^{300}$ (left) and shallow neural network (SNN) for
$SNN^{20}$, $SNN^{100}$ and $SNN^{300}$ (right). The input features are FS1+FS2.}\label{snfs1fs2}
\centering 
\begin{tabular}{c|rr|rrr} 
\hline\hline 
$m_{H^{\pm}}$  & $DNN^{100}$& $DNN^{300}$ & $SNN^{20}$& $SNN^{100}$& $SNN^{300}$ \\
\hline 
90  & 0.871 & 0.877 &0.863 & 0.874 &  0.871   \\ 
100 &0.884 & 0.881 &  0.878 & 0.881 &   0.879\\ 
110 & 0.886 & 0.881&0.888 & 0.877 &  0.869\\
120 & 0.868  &  0.868&0.897 & 0.875 &  0.843 \\[1ex] 
\hline 
\end{tabular}
\label{t2}
\end{table}
\endgroup


\section{Conclusion}
In this work we have studied signal extraction with DNNs with the particular example of charged Higgs boson search in LHC processes. We have two main findings. The first is that DNNs themselves find regions of high efficiency. In a sense,  human-engineered high-level features are offset by DNNs with similar or different combinations of the input features. Thus, DNNs are powerful tools for extracting complex features with high efficiency. The second finding is that  increasing the number of processing units in DNNs does not necessarily cause an increase in efficiency due mainly increased complexity.

The analysis here can be applied to various other processes. Light charged Higgs is a difficult signal due to its W-boson contamination but DNNs perform well in extracting the signal with our modest simulation data. A process of similar difficulty would be light scalar top quark search, which is complicated by top quark contribution. DNNs can be of critical help in extracting scalar top signal. 


\section*{Acknowledgement}
This work is supported by the T{\"U}B{\.I}TAK grant 113F167. 

\bibliographystyle{JHEP}

\bibliography{template}

\end{document}